\documentclass[aps,twocolumn,prl,superscriptaddress,amsmath,showpacs,tightenlines]{revtex4}
\usepackage{epsfig,graphicx,times}
\usepackage{amstext}
\usepackage{amsmath}            %serve per le subequazioni
\usepackage{amssymb}            %serve per il simbolo "marchio registrato", \circledR
\usepackage{graphicx}           %serve per le figure eps, ps etcyy
\usepackage{latexsym}
\usepackage{bm}

\begin{document}
\title{Generation of nonclassical photon states using a superconducting qubit in a microcavity}

\author{ Yu-xi Liu}
\affiliation{Frontier Research System,  The Institute of Physical
and Chemical Research (RIKEN), Wako-shi 351-0198, Japan}
\author{L.F. Wei}
\affiliation{Frontier Research System,  The Institute of Physical
and Chemical Research (RIKEN), Wako-shi 351-0198, Japan}
\affiliation{Institute of Quantum optics and Quantum information,
Department of Physics, Shanghai Jiaotong University, Shanghai
200030, P.R. China }
\author{Franco Nori}
\affiliation{Frontier Research System,  The Institute of Physical
and Chemical Research (RIKEN), Wako-shi 351-0198, Japan}
\affiliation{Center for Theoretical Physics, Physics Department,
Center for the Study of Complex Systems, The University of
Michigan, Ann Arbor, Michigan 48109-1120}

\date{\today}

\begin{abstract}
Based on the interaction between the radiation field and
a superconductor, we propose a way to engineer quantum states
using a SQUID charge qubit inside a microcavity. This device can
act as a deterministic single photon source as well as  generate any
Fock states and an arbitrary superposition of Fock states
for the cavity field. The controllable interaction between the
cavity field and the qubit can be realized by the tunable gate
voltage and classical magnetic field applied to the SQUID.

\pacs{42.50.Dv, 74.50.+r, 42.50.Ct}
\end{abstract}

\maketitle \pagenumbering{arabic}

The generation  of quantum states of the radiation field has been
a topic of growing interest in recent years. This is because of
possible applications in quantum communication and information
processing, such as quantum networks, secure quantum
communications, and quantum cryptography~\cite{zhuang}. Based on
the interaction between the radiation field and atoms, many
theoretical schemes have been proposed for the generation of Fock
states~\cite{fock,wl} and their arbitrary
superpositions~\cite{sahin,law}. Experiments have
  generated  single-photon states in quantum
dots~\cite{exp1}, atoms inside a microcavity~\cite{exp2}, and
other systems~\cite{exp3}. A superposition of the vacuum and
one-photon states can also be experimentally created by truncating
an input coherent state or using cavity quantum
electrodynamics~\cite{lvo}. However, how to generate an arbitrary
photon state by virtue of the interaction between the radiation
field and solid state quantum devices seems to be unknown  both
theoretically and experimentally. Recent progress in
superconducting quantum devices (e.g.,~\cite{moo,Y.Nakamura}) makes it
possible to do quantum state engineering experiments in these
systems, and also there have been proposals on superconducting
qubits interacting with the nonclassical electromagnetic
field~\cite{saidi,zhu,you,you1,gir,vourdas}.

Here, we present an experimentally feasible scheme to generate
quantum states of a single-mode cavity field in the microwave
regime by using the photon transition between the ground and first
excited states of a macroscopic two-level system formed by a
superconducting quantum interference device (SQUID). This
artificial two-level ``atom'' can be easily controlled by an
applied gate voltage $V_{g}$ and the flux $\Phi_{c}$ generated by
the classical magnetic field through the SQUID (e.g.,
~\cite{maklin,you}). The process of generating photon states in
this device includes three main steps: (i) The artificial atom
operates at the degeneracy point by choosing appropriate values
for $V_{\rm g}$ and $\Phi_{\rm c}$. There is no interaction
between the quantized cavity field and ``atom'' at this stage.
(ii) Afterwards new $V_{\rm g}$ and $\Phi_{c}$ are selected such
that the cavity field interacts resonantly with the ``atom''  and
evolves during a designated time. (iii) The above two steps can be
repeated until a desired state is obtained. Finally, the flux
$\Phi_{c}$ can be adjusted to a special value, then the
interaction is switched off, and the desired photon state appears
in the cavity. This process is similar to that of a
micromaser~\cite{fock} and it is described below.

\begin{figure}
\includegraphics[bb=117 354 629 670, width=5.0 cm, clip]{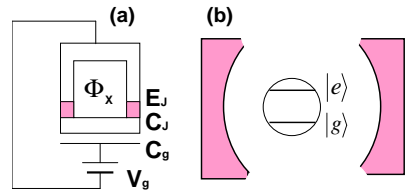}
\caption[]{(a) A charge qubit formed by a SQUID device, equivalent
to a controllable macroscopic two-level system, is placed into a
superconducting microwave cavity in (b). The coupling
between the quantized cavity field and qubit system is
realized via the magnetic flux $\Phi_{X}$ through the SQUID.}\label{fig1}
\end{figure}

{\it Model.---\,\,}The macroscopic two-level system studied here
is shown in Fig.~\ref{fig1} (a). A SQUID-type superconducting box
with $n_{\rm c}$ excess Cooper-pair charges is connected to a
superconducting loop via two identical Josephson junctions with
capacitors $C_{\rm J}$ and coupling energies $E_{\rm J}$. A
controllable gate voltage $V_{\rm g}$ is coupled to the box via a
gate capacitor $C_{\rm g}$. We assume that the superconducting
energy gap $\Delta$ is the largest energy. Then, at low
temperatures,  the quasi-particle tunneling is suppressed and no
quasi-particle excitation can be found on the island. Only Cooper
pairs coherently tunnel in the superconducting junctions. The
above assumptions are consistent with most experiments on charge
qubits. Then the standard Hamiltonian~\cite{maklin} is
\begin{eqnarray}\label{eq:h}
&&H_{\rm qb}=4E_{\rm ch}(n_{c}-n_{\rm g})^2-2E_{\rm J}\cos\!\left
(\frac{\pi\Phi_{X}}{\Phi_{0}}\right)\cos\Theta,
\end{eqnarray}
where  $\Phi_{X}$ is the total flux through the SQUID loop and
$\Phi_{0}$  the flux quantum. Thus, the superconducting loop is
used to control the Josephson coupling energy by adjusting  the
flux through this loop. Below, we show that it can also switch on
and off the qubit-field interaction. The dimensionless gate
charge, $n_{\rm g}=C_{\rm g}V_{\rm g}/2e$, is controlled by
$V_{\rm g}$. The single-electron charging energy is $E_{\rm
ch}=e^2/2(C_{\rm g}+2C_{\rm J})$. $\Theta=(\phi_{1}+\phi_{2})/2$
is the quantum mechanical conjugate of the number operator $n_{\rm
c}$ of the Cooper pairs on the box, where $\phi_{i}$ ($i=1,2$) is
the phase difference for each junction. The superconducting box is
assumed to work in the charging regime with condition $k_{B}T \ll
E_{\rm J}\ll E_{\rm ch}\ll \Delta$ where $T$ and $k_{B}$ are
temperature and Boltzmann constant respectively. If  the gate
voltages is near a degeneracy point $n_{\rm g}=1/2$, the
superconducting box is a charge qubit~\cite{maklin}, which is a
controllable two-level system characterized by the two lowest
charge states $|g\rangle$ (for $n_{\rm c}=0$) and $|e\rangle$ (for
$n_{\rm c}=1$). However, if the quasi-particle excitation cannot
be completely suppressed, a continuum of low-lying quasi-particle
states will be present, and the Hamiltonian~(\ref{eq:h}) cannot be
reduced to a system with two energy levels even when the gate
voltage is near the degeneracy point~\cite{averin}.

Now we further consider that the qubit is placed in a single-mode
microwave superconducting cavity, depicted in Fig.\,\ref{fig1}(b),
the flux $\Phi_{X}$ through the SQUID can be expressed as
~\cite{saidi,zhu, you} $\Phi_{X}=\Phi_{\rm c} +\Phi_{\rm q}$ where
the flux $\Phi_{c}$ and $\Phi_{q}=\eta \,a+\eta^{*}\,a^{\dagger}$
are generated by a classical applied magnetic field and the
quantized cavity field, respectively. Here $\eta=\int_{S}
\mathbf{u}(\mathbf{r})\cdot d\mathbf{s}$ and
$\mathbf{u}(\mathbf{r})$ is the mode function of the cavity field,
with annihilation (creation) operators $a \,(a^{\dagger})$, and
$S$ is the surface defined by the contour of the SQUID.
Considering the above, we obtain
\begin{eqnarray}\label{eq:2}
&&H=\hbar\omega a^{\dagger}a+E_{z}\sigma_{z}\\
&&-\,E_{\rm J}(\sigma_{+}+\sigma_{-})
\cos\left[\frac{\pi}{\Phi_{0}}\left(\Phi_{\rm c} I+\eta
\,a+\eta^{*}\,a^{\dagger}\right)\right]\nonumber
\end{eqnarray}
where the first two terms represent the free Hamiltonians of the
cavity field with frequency $\omega=4E_{\rm ch}/\hbar$ and the
qubit with the energy $E_{z}=-2E_{\rm ch}(1-2n_{\rm g})$, $I$ is
the identity operator. The third  term  is the nonlinear
photon-qubit interaction which is switchable by the flux
$\Phi_{c}$. The charge excited state $|e\rangle$ and ground state
$|g\rangle$ correspond to the eigenstates $|\!\downarrow\rangle$
and $|\!\uparrow\rangle$ of the spin operator $\sigma_{z}$,
respectively.  The cosine in Eq.~(\ref{eq:2}) can be further
decomposed into classical and quantized parts, and the quantized
parts $\sin[\pi(\eta\, a+H.c.)/\Phi_{0}]$ and $\cos[\pi(\eta\,
a+H.c.)/\Phi_{0}]$ can be further expanded as a power series in $a
\,(a^{\dagger})$. Here, the single photon transition between the
states $|e,n\rangle$ and $|g, n+1\rangle$ satisfies the condition
$(\pi|\eta|/\Phi_{0})\sqrt{n+1}\ll 1$, where $n$ is the number of
photons; therefore all higher orders of $\pi|\eta|/\Phi_{0}$ can
be neglected and only a single-photon transition is kept in the
expansion of Eq.~(\ref{eq:2}). Using the notation for  trapped ion
systems (e.g., ~\cite{wei}),  the first red  (blue) sideband
excitations  $\beta a\sigma_{+}+H.c.$ $(\beta a\sigma_{-}+H.c.)$
for interactions of the cavity field and the qubit~\cite{zhu},
with photon-qubit coupling constant $\beta=(\pi\eta E_{\rm
J}/\Phi_{0})\sin(\pi\Phi_{c}/\Phi_{0})$, can be obtained by
adjusting the gate voltages $V_{g}$ and the flux $\Phi_{c}$. They
correspond to $2E_{z}=\hbar\omega\, (2E_{z}=-\hbar\omega)$ and
dimensionless gate charge $n_{\rm g}=1$ ($n_{\rm g}=0$). Also
$\xi(\sigma_{+}+\sigma_{-})$ with $\xi= E_{\rm
J}\cos(\pi\Phi_{c}/\Phi_{0})$ is called the carrier~\cite{zhu},
which corresponds to $n_{\rm g}=1/2$. The Hamiltonian
(\ref{eq:2}), with the above assumptions, is our model.

{\it Preparation process.---\,\,\,}We choose $|0,g\rangle$  as our
initial state, where the cavity field is in the vacuum state
$|0\rangle$ and the qubit is in the ground state $|g\rangle$. The goal
is to prepare an arbitrary pure state of the cavity field
\begin{equation}\label{eq:3}
|\psi\rangle=\sum_{n=0}^{N}c_{n}|n,g\rangle=|g\rangle\otimes\sum_{n=0}^{N}c_{n}|n\rangle
\end{equation}
where $|n\rangle$ denotes the Fock states of the cavity field with
excitation number $n=0,1,2,\cdots$. A Fock state $|m\rangle$ with
$m$ photons is a special case of Eq.~(\ref{eq:3}) with conditions
$c_{n}=0$ for all $n\neq m$ with $0<m\leq N$.

Thermal photons in the cavity have to
be suppressed in order to obtain the vacuum state $|0\rangle$. In
the microwave region $0.1 \sim 15$ cm, the mean number of thermal
photons $\langle n_{\rm th}\rangle$ satisfies $ 3.0 \times
10^{-208}\leq \langle n_{\rm th}\rangle\leq 0.043$ at $T=30$ mK,
and $ 1.7\times 10^{-104}\leq \langle n_{\rm th}\rangle\leq 0.26$
at $T=60$ mK. These temperatures can be  obtained experimentally
(e.g., in~\cite{Y.Nakamura,tsai}).

After the  system is initialized, two different processes are
required to engineer the state of the cavity field. The first
process involves rotating the qubit state, but keeping  the cavity
field state unchanged. This stage can be experimentally realized
by tuning the gate voltage and classical magnetic field such that
$n_{\rm g}=1/2$ and $\Phi_{\rm  c}=0$; then the time evolution
operator $U_{\rm  C}(t)$ of the qubit in the interaction picture
is
\begin{equation}\label{eq:5}
U_{\rm C}(t)= \cos(\Omega_{1} t)I+i\sin(\Omega_{1}
t)(|g\rangle\langle e|+|e\rangle\langle g|)
\end{equation}
where $\Omega_{1} =E_{\rm J}/\hbar$. The subscript  ``C''  in
$U_{\rm C}(t)$ denotes the carrier process, which can superpose
two levels of the qubit, and it can also flip the ground state
$|g\rangle$ or excited state $|e\rangle$ to each other, after a
time $t=\pi(2p-1)/2\Omega_{1}$, with positive integer $p$.

The second process is the first red  (blue) sideband excitation,
which can be realized by tuning the gate voltage and classical
magnetic field such that $n_{\rm g}=1$ ($n_{\rm g}=0$) and
$\Phi_{c}=\Phi_{0}/2$. Thus, in the interaction picture,  the time
evolution operators $U_{\rm R}(t)$ for the red ($U_{\rm B}(t)$ for
the blue) of the cavity field and qubit can be
expressed~\cite{sst} as
\begin{eqnarray}\label{eq:7}
U_{\rm R}(t)&=&R_{ee}(t)|e\rangle\langle
e|+R_{gg}(t)|g\rangle\langle
g|\nonumber\\
&-&iR_{ge}(t)|g\rangle\langle e|-iR_{eg}(t)|e\rangle\langle g|
\end{eqnarray}
or
\begin{eqnarray}\label{eq:11}
U_{\rm B}(t)&=&R_{gg}(t)|e\rangle\langle
e|+R_{ee}(t)|g\rangle\langle g|\nonumber\\
&-& iR_{ge}(t)|e\rangle\langle g|-iR_{eg}(t)|g\rangle\langle e|
\end{eqnarray}
with $ R_{eg}(t)=\left[e^{i\theta}\sin\left(
|\Omega_{2}|t\sqrt{aa^{\dagger}}\right)/\sqrt{aa^{\dagger}}\right]
a$, $R_{ge}(t)=B^{\dagger}_{eg}(t)$,
$R_{ee}(t)=\cos\left(|\Omega_{2}|t\sqrt{aa^{\dagger}}\right)$, and
$R_{gg}(t)=\cos\left(|\Omega_{2}|t\sqrt{a^{\dagger}a}\right)$,
where we have assumed that $\Omega_{2}=\pi \eta E_{\rm
J}/\hbar\Phi_{0}=|\Omega_{2}|e^{i\theta}$, in which the phase
$\theta$ depends on the mode function of the cavity field
$u(\mathbf{r})$. The red sideband excitation described by operator
$U_{R}(t)$ can entangle $|g,n+1\rangle$ with $|e,n\rangle$, or
flip $|g,n+1\rangle$ to $|e,n\rangle$ and vice versa,  by choosing
the duration of the interaction between the cavity field and the
qubit. From Eq.~(\ref{eq:7}), it is easy to verify that the
emission probability $P_{g}$ of the upper level for the qubit is
$P_{g}=\sin^{2}(|\Omega_{2}|t\sqrt{n+1})$. We find that $P_{g}=1$
when $|\Omega_{2}|t\sqrt{n+1}=\pi(2k-1)/2$,  with positive integer
$k$. So when $t=\pi(2k-1)/(2|\Omega_{2}|\sqrt{n+1})$, there are
$n+1$ photons in the cavity and the qubit is in its ground state.
The first blue sideband excitation,  denoted by $U_{\rm B}(t)$,
can entangle state $|e, n+1\rangle$ with state $|g, n\rangle$, or
flip $|e, n+1\rangle$  to $|g, n\rangle$ and vice versa. Below we
use the carrier and the first red sideband excitation, represented
by $U_{\rm C}(t)$ and $U_{\rm R}(t)$,  as an example showing the
generation of an arbitrary quantum state of the cavity field.

Using the quantum operations $U_{\rm C}(t)$ and $U_{\rm R}(t)$ in
Eqs.~(\ref{eq:5}) and (\ref{eq:7}), the single photon state
$|1\rangle$ can be generated from the initial vacuum state
$|0\rangle$. That is, we can first flip the ground state of the
qubit to the excited state when the condition
$\Omega_{1}t_{1}=\pi/2$ is satisfied for the carrier $U_{\rm
C}(t_{1})$, then we turn on the first red sideband excitation
$U_{\rm R}(t_{2})$ and let the photon-qubit system evolve a time
$t_{2}$ satisfying the condition $|\Omega_{2}|t_{2}=\pi/2$.
Finally, we adjust the classical magnetic field such that
$\Phi_{\rm c}=0$; thus the interaction between the cavity field
and qubit vanishes, and a single-photon state exists in the
cavity, that is,
\begin{equation}
|1\rangle\otimes |g\rangle =U_{\rm R}(t_{2}|)\,U_{\rm
C}(t_{1})\,|0\rangle\otimes |g\rangle.
\end{equation}
Also any Fock state $|m\rangle$ can  be easily created from the
vacuum state $|0\rangle$ by alternatively turning on and off
the quantum operations in Eqs.~(\ref{eq:5}-\ref{eq:7}) to
 excite the qubit and emit photons during the time
interval $T$. The latter is divided by $2m$ subintervals $\tau_{1}, \, \tau_{2},
\,\cdots,\,\tau_{2l-1}, \,\tau_{2l} \,\cdots,\, \tau_{2m-1}, \,
\tau_{2m}$
which satisfy conditions $|\Omega_{1}|\tau_{2l-1}=\pi/2$ and
$|\Omega_{2}|\tau_{2l}\sqrt{l+1}=\pi/2$ where $l=1,\cdots,m$. This
process can be described as
\begin{equation}
|m\rangle\otimes |g\rangle=U_{\rm R}(\tau_{2m})U_{C}(
\tau_{2m-1})\cdots U_{\rm
R}(\tau_{2})U_{C}(\tau_{1})|0\rangle\otimes |g\rangle.
\end{equation}
Finally, the classical magnetic field is changed such that
$\Phi_{\rm c}=\Phi_{0}$, and an $n$-photon state is provided in
the cavity.

Our next goal is to prepare superpositions of different Fock
states (e.g., $\alpha_{1}|0\rangle+\alpha_{2}|1\rangle$) for the
vacuum $|0\rangle$ and single photon $|1\rangle$ states. This very
important state can be deterministically generated by two steps,
$U_{\rm C}(t'_{1})$ and $U_{\rm R}(t'_{2})$,  with
$t'_{2}=\pi/2|\Omega_{2}|$; that is
\begin{equation}
(\alpha_{1}|0\rangle+\alpha_{2}|1\rangle)\otimes|g\rangle=U_{\rm
R}(t^{\prime}_{2})U_{\rm C}(t^{\prime}_{1})
|0\rangle\otimes|g\rangle
\end{equation}
where the operation time $t'_{1}$ determines the weights of the
coefficients of the superposition
$\alpha_{1}=\cos(\Omega_{1}t'_{1})$ and
$\alpha_{2}=e^{-i\theta}\sin(\Omega_{1}t'_{1})$. If the condition
$t'_{1}=\pi/4\Omega_{1}$ is satisfied, then we have a superposition
$(|0\rangle+e^{-i\theta}|1\rangle)/\sqrt{2}$ with equal
probabilities for each component and the relative phase between
them can be further specified by the phase of the mode of the cavity field.

An arbitrary target state (\ref{eq:3}) can be generated from the
initial state by alternatively switching on and off the carrier
and first red sideband excitation during the time $T'$, which can
be divided into $2n$ subintervals $\tau'_{1},\cdots, \tau'_{2n}$. That is,
the target state can be deterministically generated as follows
\begin{equation}\label{eq:9}
|\psi\rangle=\sum_{n=0}^{N}c_{n}|n, g\rangle=U(T')|0, g\rangle,
\end{equation}
where $U(T')$ is determined by a sequence of time evolution
operators associated with chosen time subintervals as $
U(T')=U_{\rm R}(\tau'_{2n})U_{\rm C}(\tau'_{2n-1})\cdots U_{\rm
R}(\tau'_{2})U_{\rm C}(\tau'_{1})$. Therefore, the coefficients
$c_{n}$ are
\begin{equation}
c_{n}=\langle g, n|U_{\rm R}(\tau'_{2n})U_{\rm
C}(\tau'_{2n-1})\cdots U_{\rm R}(\tau'_{2})U_{\rm C}(\tau'_{1})|0,
g\rangle.
\end{equation}
Reference~\cite{law} has explicitly discussed how to adjust the
rescaled times to obtain the expected state by solving the inverse
evolution of Eq.~(\ref{eq:9}). Ideally, any state of the cavity
field can be created according to our proposal by adjusting the
gate voltage, classical magnetic field,  and duration of the
photon-qubit interaction. It is very easy to check that the state
(\ref{eq:3}) can also be created by the carrier and blue sideband
excitation whose time evolutions $U_{\rm C}(t)$ and $U_{\rm B}(t)$
are described by Eqs. (\ref{eq:5}) and (\ref{eq:11}).

{\it Environmental effects.---\,\,\,}We now discuss the
environmental effects on the prepared states, which are
actually limited by the following time scales: the relaxation time
$T_{1}$, the preparation time $\tau_{e}$ of the excited state, and
the dephasing time $T_{2}$ of the qubit,  the lifetime $\tau_{p}$
of the photon and an effective interaction time $\tau^{(n)}_{c}$ which
corresponds to the transition from $|n,e\rangle$ and
$|n+1,g\rangle$. If $T_{1}, \,\, \tau_{p}\gg \tau_{e},\, \,
\tau^{(n)}_{c}$, then the Fock states can be prepared. If the condition
$T_{1}, \,\,T_{2}, \,\, \tau_{p}\gg \tau_{e},\, \, \tau^{(n)}_{c}$ is
satisfied, then the superposition can also be obtained.

\begin{figure}
\includegraphics[width=6.0cm]{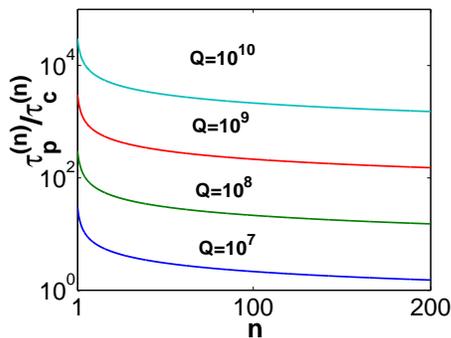}
\caption[]{Ratio $\tau^{(n)}_{p}/\tau^{(n)}_{c}$, versus photon number $n\geq1$,
of the lifetime $\tau^{(n)}_{p}$ of the photon number state $|n\rangle$ and the effective operation
time $\tau^{(n)}_{c}$. The latter corresponds to the
transition from $|n,e\rangle$ to $|n+1,g\rangle$ for
a $10\mu$m $\times 10\mu$m SQUID in the full-wave cavity.}\label{fig2}
\end{figure}

Now let us estimate the photon number of the obtainable Fock state
in a full-wave cavity. In microwave experiments, it is possible to
obtain very high-$Q$ superconducting cavities, with $Q$  values
around $3\times 10^{8}$ to $5\times10^{10}$~\cite{fock,life},
which correspond to the lifetimes of the microwave region  from
$0.001\leq\tau_{p}\leq 0.15$ s to $0.167\leq\tau_{p}\leq 25$ s.
The parameters of the charge qubit~\cite{kw} without the cavity
are $2E_{\rm J}/h=13.0$ GHz (so the operation time corresponding
to a completely excited qubit is about $\tau_{e}\approx 3.8\times
10^{-11}$ s). The lifetime of the excited-state for the qubit
$T_{1}=1.3\times 10^{-6}$ s, i.e. $\tau_{e}\ll T_{1}$. For an
estimate of the interaction coupling between the cavity field and
the qubit, we assume that the cavity mode function is taken as a
standing-wave form  such as
$B_{x}=-i\sqrt{\hbar\omega/\varepsilon_{0}V
c^{2}}(a-a^{\dagger})\cos(k z)$, with polarization along the
normal direction of the surface area of the SQUID,  located at an
antinode of the standing-wave mode; then the interaction between
the cavity field and the qubit reaches its maximum and the
interaction strength can be expressed as $|\beta|=\pi|\eta|E_{\rm
J}/\Phi_{0}=(\pi S E_{\rm
J}/c\Phi_{0})\sqrt{\hbar\omega/\varepsilon_{0}V}$. For example, if
the wavelength of the cavity mode is taken as $\lambda_{1}=0.1$
cm, then $\pi |\eta|/\Phi_{0}\approx 7.38\times 10^{-5}\ll 1$,
where the dimension of the SQUID is taken as $10\,\mu$m and the
mode function $u(\mathbf{r})$ is assumed to be independent of the
integral area because the dimension of the SQUID, $10 \,\,\mu$m,
is much less than $0.1$ cm, for the wavelength of the cavity mode.
In this case, $\tau^{(0)}_{c}\approx 5.0 \times 10^{-7} {\rm s}$,
which is less than one order of magnitude of the excited lifetime
$T_{1}$. This means that the qubit in its excited state can emit a
photon before it relaxes to its ground state. But if we take the
dimension of the SQUID as $1 \,\mu$m, the coupling between the
cavity field and the qubit is two orders of magnitude smaller than
for the $10\, \mu$m SQUID, and then the interaction time is
$5.0\times 10^{-5}>T_{1}$. Therefore, in this case, the qubit
relaxes to the ground state before the photon can be emitted from
the qubit, and thus it is difficult to obtain a photon state. In
Fig.~(\ref{fig2}), we plot the ratio
$\tau^{(n)}_{p}/\tau^{(n)}_{c}$ between the
lifetime~\cite{lifetime} $\tau^{(n)}_{p}=\tau_{p}/n$ (here,
$\tau^{(1)}_{p}=\tau_{p}$) of the Fock state $|n\rangle$ for the
zero-temperature environment and the effective operation time
$\tau^{(n)}_{p}$ of transitions from state $|n,e\rangle$ to
$|n+1,g\rangle$ for different values of $Q$ and for $\lambda=0.1
\,{\rm cm}$.  Fig.~\ref{fig2} shows that the photon number of the
prepared Fock states can reach $10^{2}$ in the above mentioned
high-$Q$ cavity. But if the $Q$ values are less than $10^{7}$, it
might be difficult to prepare a photon state with our estimated
coupling. We  also find that a longer microwave in the full-wave
cavity corresponds to a longer $\tau^{(n)}_{c}$ for a fixed Q,
which means that it is easy to create photon states for shorter
microwaves. For example, if the wavelength is taken as $\lambda=1$
cm, then the coupling between the qubit-photon in the full-wave
cavity might not be strong enough for generating Fock states
within the currently known experimental data for $T_{1}$. So for
longer microwaves, we can make a smaller cavity and place the
qubit where the  qubit-photon interaction is maximum.

If we want to prepare a superposition of  different Fock states of
the cavity field, we need to consider $T_{2}$, which is of the
order of a few ns (e.g., $5$ ns in ~\cite{Y.Nakamura}). Then the
survival time of the entangled state between the cavity field and
qubit, which is required for the preparation of the superposition
of the Fock states, maybe be very short. With the improvement of read-out
techniques, a longer dephasing time can make our proposal far more realizable.

{\it Discussions.---\,\,\,}We propose a scheme for deterministically
generating  nonclassical photon states via the interaction of
photons and a charge qubit. Indeed, the Fock state can be
prepared with current technology. The superposition would be
easier to obtain by increasing the dephasing time $T_{2}$ and the
qubit-photon coupling strength. Our discussions above are based on
 experimental values for $T_{1}$ and $T_{2}$ without the cavity; the
decoherence may  become shorter when the SQUID is placed inside
the cavity. Further, in order to obtain a stronger coupling, the
following steps would help to increase the qubit-field coupling
strength: i) decrease the volume $V$ of the cavity; ii) increase
the area $S$ of the SQUID; iii) increase the Josephson coupling
energy $E_{\rm J}$ under the condition $E_{\rm J}\ll E_{\rm ch}$.
We can also put a high permeability $\mu$ material inside the
SQUID loop~\cite{you}, then the qubit-field coupling strength can
increase to $\mu|\beta|$, because the relative permeability in
ferromagnetic materials can be $10^{2}$ to $10^{6}$, and might
partly compensate some of the decoherence effects due to the $\mu$
material itself. Increasing the SQUID dimension and decreasing the
cavity volume will reduce the maximum allowed number of SQUIDs
inside the cavity making it unadvantageous for quantum computing.
However, one qubit is enough for the generation of nonclassical
photon states, our goal here. We note that Girvin {\it et
al.}~\cite{gir} proposed a different system in which the coupling
of the photon-qubit can reach $10^{8}$ Hz,  corresponding to $
\tau^{(0)}_{c}\sim 10^{-9}$ s. We are considering how to generate
nonclassical photon states by using such a system. This scheme
might not be easy to generalize in a straightforward manner to the
flux qubit case. This because the interaction between the
flux-qubit and the cavity field cannot be switched on and off in
the same way for the charge qubit. However in some modified
manner, it should be possible to generalize this scheme.

We thank X. Hu helpful comments. This work was supported in part
by the US NSA and ARDA under AFOSR contract No. F49620-02-1-0334,
and by the NSF grant No. EIA-0130383.


\begin{thebibliography}{99}
\bibitem{zhuang}M.N. Nielsen and I.L. Chuang, {\it Quantum Computation and Quantum
Information} (Cambridge University Press 2000).


\bibitem{fock}J. Krause, {\it et al.}, Phys. Rev. A {\bf 36}, 4547 (1987);
{\it ibid.} {\bf 39}, 1915 (1989).

\bibitem{wl} W. Leonski, {\it et al.}, Phys. Rev. A
{\bf 49}, R20 (1994);  W. Leonski, {\it ibid.} {\bf 54}, 3369
(1996); M. Brune, {\it et al.}, Phys. Rev. Lett. {\bf 65}, 976
(1990); M. J. Holland, {\it et al.}, {\it ibid}. {\bf 67}, 1716
(1991).


\bibitem{law}C.K. Law, {\it et al.}, Phys. Rev. Lett. {\bf
76}, 1055 (1996).

\bibitem{sahin}K. Vogel, {\it et al.}, Phys. Rev. Lett. {\bf 71}, 1816 (1993);
A.S. Parkins, {\it et al.}, {\it ibid.} {\bf 71}, 3095 (1993);
Phys. Rev. A {\bf 51}, 1578 (1995); D.T. Pegg, {\it et al.}, Phys.
Rev. Lett. {\bf 81}, 1604 (1998).

\bibitem{exp1}P. Michler, {\it et al.}, Science {\bf 290},
2282 (2000); J. Kim, {\it et al.}, Nature {\bf 397}, 500 (1999);
P. Michler, {\it et al.}, {\it ibid.} {\bf 406}, 968 (2000).

\bibitem{exp2}C.J. Hood, {\it et al.}, Science {\bf 287}, 1447 (2000);
B.T.H. Varcoe, {\it et al.}, Nature {\bf 403}, 743 (2000); P.W.H.
Pinkse, {\it et al.}, {\it ibid.} {\bf 404}, 365 (2000); P.
Bertet, {\it et al.}, Phys. Rev. Lett. {\bf 88}, 143601 (2002); A.
Kuhn, {\it et al.}, {\it ibid.} {\bf 89}, 067901 (2002).


\bibitem{exp3}B. Lounis {\it et al.}, Nature {\bf 407},
491 (2000);  C.K. Hong, {\it et al.}, Phys. Rev. Lett. {\bf 56},
58 (1986); C. Brunel, {\it et al.}, {\it ibid.} {\bf 83}, 2722
(1999); C. Kurtsiefer, {\it et al.}, {\it ibid.} {\bf 85}, 290
(2000); C. Santori, {\it et al.}, {\it ibid.} {\bf 86}, 1502
(2001).%; F. Treussart, {\it et al.}, {\it ibid.} {\bf 89}, 093601
%(2002).



\bibitem{lvo}A.I. Lvovsky, {\it et al.}, Phys. Rev. Lett. {\bf 88},
250401 (2002);  X. Maitre, {\it ibid.} {\bf 79}, 769 (1997).

\bibitem{moo}J.E. Mooij, {\it et al.}, Science {\bf 285},
1036 (1999); C.H. van der Wal, {\it et al.}, {\it ibid.} {\bf
290}, 773 (2000);  S. Han, {\it et al.}, {\it ibid.} {\bf 293},
1457 (2001); D. Vion, {\it et al.}, {\it ibid.} {\bf 296}, 886
(2002); Y. Yu, {\it et al}, {\it ibid.} {\bf 296}, 889 (2002);  I.
Chiorescu, {\it et al.}, {\it ibid.} {\bf 299}, 1869 (2003); A.J.
Berkley, {\it et al.}, {\it ibid.} {\bf 300}, 1548 (2003); J.R.
Friedman, {\it et al.}, Nature {\bf 406}, 43 (2000).

\bibitem{Y.Nakamura}Y. Nakamura, {\it et al.}, Nature {\bf 398},
786 (1999);  Y.A. Pashkin, {\it et al.}, {\it ibid.} {\bf 421},
823 (2003).


\bibitem{saidi}W.A. Al-Saidi, {\it et al.}, Phys. Rev. B {\bf
65}, 224512 (2002).

\bibitem{zhu}S.L. Zhu, {\it et.al}, Phys. Rev. A {\bf 68}, 034303 (2003).

\bibitem{you}J.Q. You and F. Nori, Phys. Rev. B {\bf 68}, 064509 (2003).

\bibitem{you1}J.Q. You, {\it et. al},  Phys. Rev. B {\bf 68}, 024510
(2003).

\bibitem{vourdas}C.P. Yang, {\it et.al}, Phys. Rev. A {\bf 67}, 042311 (2003)

\bibitem{gir}S.M. Girvin, {\it et al.}, cond-mat/0310670.

\bibitem{maklin}Y. Makhlin, {\it et al.}, Rev. Mod.
Phys. {\bf 73}, 357 (2001).

\bibitem{averin}D.V. Averin and Y.V. Nazarov, Phys. Rev. Lett. {\bf
69}, 1993 (1992); M. Tinkham, {\it Introduction to
Superconductivity} (McGraw-Hill, New York, 1996).

\bibitem{wei}D.M. Meekhof,  {\it et al.}, Phys. Rev. Lett. {\bf 76},
1796 (1996); A. Ben-Kish, {\it et al.}, {\it ibid.} {\bf 90},
037902 (2003); B. Kneer, {\it et al.}, Phys. Rev. A {\bf 57}, 2096
(1998); L.F. Wei, {\it et al.}, quant-ph/0308079.


\bibitem{tsai}Y. Nakamura, {\it et al.},  Phys.
Rev. Lett. {\bf 87}, 246601 (2001); {\it ibid.} {\bf 88}, 047901
(2002).


\bibitem{sst}S. Stenholm, Phys. Rep. {\bf 6}, 1(1973).


\bibitem{life}S. Brattke, {\it et al.}, Phys. Rev. Lett. {\bf 86}, 3534
(2001); J. M. Raimond, {\it et al.}, Rev. Mod. Phys. {\bf 73}, 565
(2001).

\bibitem{kw}K.W. Lehnert, {\it et al.}, Phys. Rev. Lett. {\bf
90}, 027002 (2003).


\bibitem{lifetime}V. Perinova and A. Luks, Phys. Rev. A {\bf 41},
414 (1990).

\end{thebibliography}
\end{document}